\documentclass[aps,prl,twocolumn,showpacs,amsmath,amsmath,amssymb]{revtex4}
\usepackage{amssymb}
\usepackage{graphicx}
\usepackage{dcolumn}
\usepackage{bm}
\usepackage{psfrag}
\usepackage{bbold}
\usepackage{float}

\begin{document}

\author{Ryan Jones}
\author{Reece Saint}
\author{Beatriz Olmos}

\affiliation{School
of Physics and Astronomy, The University of Nottingham, Nottingham,
NG7 2RD, United Kingdom}

\title{Far-field resonance fluorescence from a dipole-interacting laser-driven cold atomic gas}
\date{\today}

\begin{abstract}
We analyze the temporal response of the fluorescence light that is emitted from a dense gas of cold atoms driven by a laser. When the average interatomic distance is smaller than the wavelength of the photons scattered by the atoms, the system exhibits strong dipolar interactions and collective dissipation. We solve the exact dynamics of small systems with different geometries and show how these collective features are manifest in the scattered light properties such as the photon emission rate, the power spectrum and the second-order correlation function. By calculating these quantities beyond the weak driving limit, we make progress in understanding the signatures of collective behavior in these many-body systems. Furthermore, we clarify the role of disorder on the resonance fluorescence, of direct relevance for recent experimental efforts that aim at the exploration of many-body effects in dipole-dipole interacting gases of atoms.
\end{abstract}

\pacs{}

\maketitle

\noindent\textit{Introduction.} Strong dipole-dipole interactions are induced in a gas of emitters due to virtual exchange of photons when the average distance between the emitters is smaller than the wavelength associated to the emitted photons. In these \textit{dense} gases, the radiation properties differ drastically from the \textit{dilute} case due to the emergence of collective super- and sub-radiant emission modes. The unique character of such a system was studied for the first time decades ago in the seminal papers by Dicke, Lehmberg and Agarwal among others \cite{Dicke54,Lehmberg70,Agarwal70}.

The unprecedented experimental control available nowadays over the trapping and interactions in ultracold atomic gases \cite{Bloch08} has sparked a renewed interest in the investigation of these fundamentally collective effects. Experimental measurements of features such as the collective Lamb shift \cite{Rohlsberger10,Keaveney12,Meir14}, suppression of light scattering and modified spectra from dense samples of atoms \cite{Pellegrino14,Kwong14,Kwong15,Jennewein15,Bromley16,Jenkins16} have been recently realized. Theoretical works so far have been constrained to the study of the limit of very weak driving \cite{Fleischhauer99,Scully09,Svidzinsky10,Bienaime12,Bienaime13,Olmos13,Li13,Bettles15}, small systems of two or three atoms \cite{Kus81,James93,Hettich02,Das08,Wang10,Zoubi12} or dilute gases under strong driving conditions \cite{Ott13}. These, however, do not provide a complete picture and leave a number of unanswered questions: (A) How does the presence of strong laser driving affect the signatures of cooperativity detectable in the fluorescence photons scattered from a dipolar system? (B) Photon emission rate and excitation number have a one-to-one relation in dilute gases that is broken when the dissipation becomes collective. Can one observe this phenomenon in a dense atomic gas? (C) How does the specific external configuration of the atoms affect the previous results? I.e., are there any differences expected to arise in experimental setups with atoms in ordered (e.g. optical lattices) and disordered configurations?

\begin{figure}[t]
\includegraphics[width=\columnwidth]{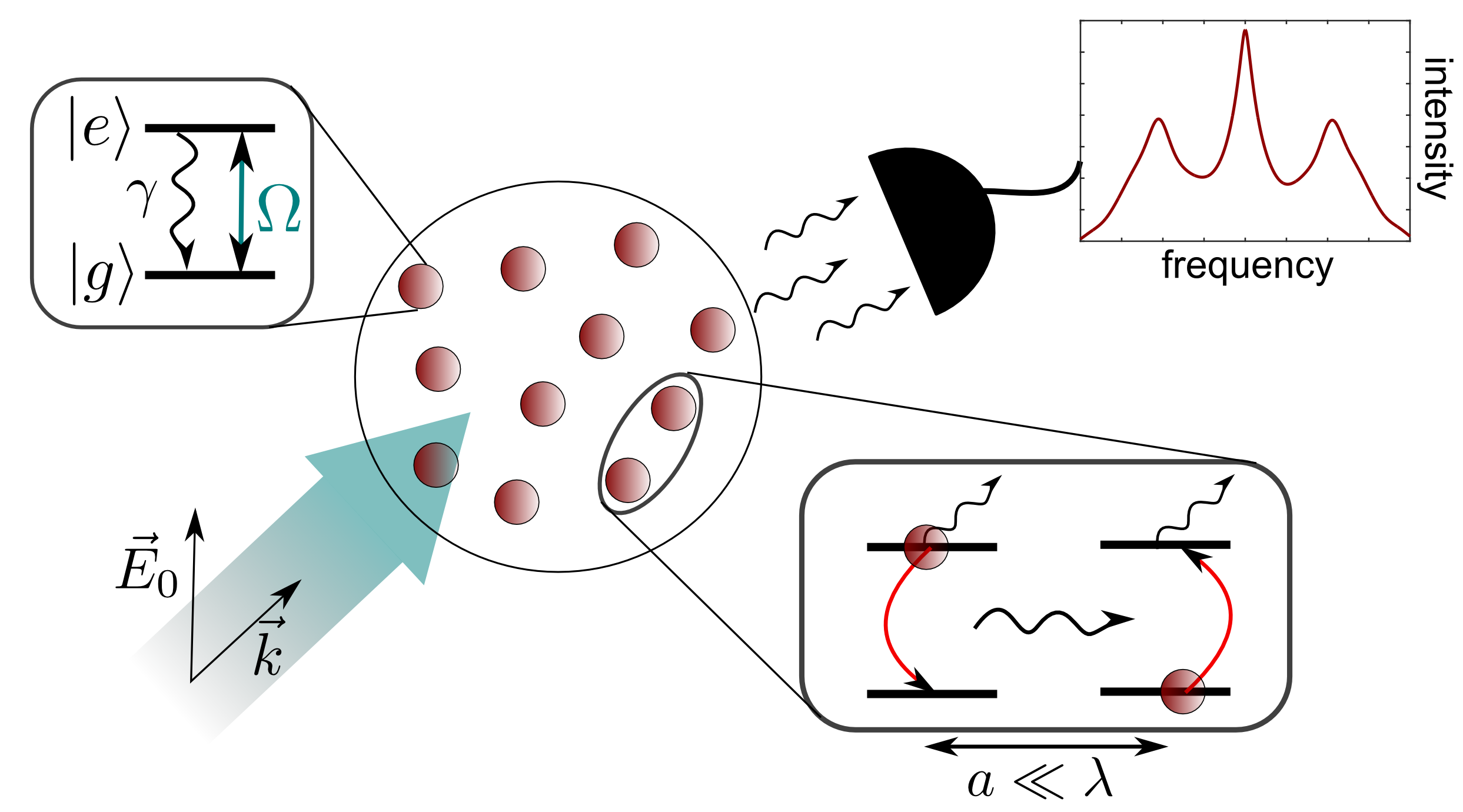}
\caption{(Color online) An ensemble of $N$ two-level atoms is illuminated uniformly by a laser field polarized along the $z$-axis, which resonantly couples the atomic $\left|g\right>-\left|e\right>$ transition with Rabi frequency $\Omega$. The virtual exchange of photons gives rise to long-range exchange interactions and collective dissipation when the interatomic distance $a$ between the atoms is smaller than the wavelength of the transition $\lambda$. A photodetector is used to detect the emitted photons and obtain the intensity and spectrum of the resonance fluorescence.}\label{Fig1}
\end{figure}

We tackle the above questions in this paper by performing a detailed theoretical analysis of the excitation number, photon emission rate, power spectrum and second-order correlations of the far-field fluorescence from a resonantly driven gas of two-level atoms in the stationary state. We solve numerically the \textit{exact} dynamics of atomic systems of up to $7$ atoms for a broad range of values of the laser driving. We show that this gives insights into the behavior of larger systems and hence make the results of this paper of direct relevance to current experimental efforts that study the effect of dipole-dipole interactions and collective dissipation in the optical response of a cold atomic system in the absence of inhomogeneous broadening \cite{Pellegrino14,Kwong14,Kwong15,Jennewein15}. In particular, in order to illustrate the important role of the external geometry in these cold atomic gases, we have analyzed the emission properties from an ordered system --a one-dimensional (1D) lattice of atoms-- and a three-dimensional (3D) disordered gas, where the positions of the atoms are chosen randomly and the results are averaged over many different realizations, as it is done experimentally.

\noindent\textit{The system.} We consider an ensemble of $N$ atoms either confined in an optical lattice or in a disordered gas. All atoms are initially assumed to be in the electronic ground state, $|g\rangle$. An external laser field linearly polarized along the $z$-axis is then applied to couple resonantly the two internal states $|g\rangle$ and $|e\rangle$. The average interatomic distance between neighboring atoms, $a$, is here considered to be much shorter than the transition wavelength $\lambda$ (see Fig.~\ref{Fig1}). As a consequence, strong long-range interactions are induced among the atoms and the photon emission acquires a marked collective character \cite{Dicke54,Lehmberg70,Agarwal70}.

The dynamics of the ensemble is described by the master equation
\begin{equation}\label{MasterEq}
\dot{\rho}=-\frac{i}{\hbar}\left[H,\rho\right]+{\cal{D}}(\rho),
\end{equation}
where $\rho$ is the atomic density matrix. The many-body Hamiltonian $H$, which describes the coherent time-evolution of this open quantum system, is expressed as
\begin{equation*}
H=\hbar\sum_{\alpha=1}^N\Omega\left( b^\dag_{\alpha}e^{i{\bf{k}}\cdot{\bf{r}}_{\alpha}} +b_{\alpha}e^{-i{\bf{k}}\cdot{\bf{r}}_{\alpha}}\right) +\hbar\sum_{\alpha\neq\beta}V_{\alpha\beta}b^\dag_{\alpha}b_{\beta}.
\end{equation*}
Here we have defined the atomic transition operator $b_{\alpha}\equiv|g\rangle_\alpha\!\langle e|$ for the $\alpha$-th atom. The atom-laser coupling strength is given by the Rabi frequency $\Omega=dE_{0}/(2\hbar)$ with $E_{0}$ being the amplitude of the external homogeneous laser field and $d$ the transition dipole moment. The driving field wavevector is denoted by ${\bf{k}}=k\mathbf{\hat{y}}$ (see Fig.~\ref{Fig1}) and the spatial position of the $\alpha$-th atom is ${\bf{r}}_{\alpha}$. The long-range coherent interaction between the $\alpha$-th and $\beta$-th atoms separated by $\mathbf{r}_{\alpha\beta}=r_{\alpha\beta}\hat{\mathbf{r}}_{\alpha\beta}$ is characterized by the coefficient matrix
\begin{equation*}
V_{\alpha\beta}=\frac{3\gamma}{4}\left[y_0(\kappa_{\alpha\beta})-\frac{y_1(\kappa_{\alpha\beta})}{\kappa_{\alpha\beta}} +y_2(\kappa_{\alpha\beta})(\hat{\mathbf{d}}\cdot\hat{\mathbf{r}}_{\alpha\beta})^2\right]
\end{equation*}
where $y_n(x)$ denotes the spherical Bessel function of second kind, $\gamma$ the spontaneous decay rate of the excited state, $\hat{\mathbf{d}}$ the direction of the transition dipole moment and where we have defined the reduced length $\kappa_{\alpha\beta}=2\pi r_{\alpha\beta}/\lambda$. The second term of Eq.~(\ref{MasterEq}) describes the spontaneous emission of photons from the system and takes the form
\begin{equation*}
{\cal{D}}(\rho)=\sum_{\alpha,\beta}\Gamma_{\alpha\beta}\left(b_{\alpha}\rho b^\dag_{\beta}-\frac{1}{2}\left\{b^\dag_{\alpha}b_{\beta},\rho\right\}\right),
\end{equation*}
where
\begin{equation*}
\Gamma_{\alpha\beta}=\frac{3\gamma}{2}\left[j_0(\kappa_{\alpha\beta})-\frac{j_1(\kappa_{\alpha\beta})}{\kappa_{\alpha\beta}} +j_2(\kappa_{\alpha\beta})(\hat{\mathbf{d}}\cdot\hat{\mathbf{r}}_{\alpha\beta})^2\right]
\end{equation*}
represents the strength of dissipative coupling between two atoms. Here, $j_n(x)$ denotes the spherical Bessel function of first kind.

We are interested in the regime where the distance between the atoms is smaller or comparable to $\lambda$, i.e. $\kappa_{\alpha\beta}\leq1$. One can unravel the collective character of the dissipation in this regime by rewriting the dissipation term in diagonal form as
\begin{equation*}
{\cal{D}}(\rho)=\sum_{m=1}^{N}\gamma_{m}\left(J_{m}\rho J^\dag_{m}-\frac{1}{2}\left\{J^\dag_{m}J_{m},\rho\right\}\right).
\end{equation*}
In this form, it is easy to identify $J_{m}=\sum_{\alpha}M_{m\alpha}b_{\alpha}$ (with $M$ containing the eigenvectors of the matrix $\Gamma$) as an operator associated to the emission of a photon and $\gamma_{m}=\sum_{\alpha,\beta}M_{m\alpha}\Gamma_{\alpha\beta}M^\dag_{\beta m}$ as the rate at which such an emission takes place. The structure of these operators dictates to which extent the emission and the atomic excitation are coupled, as it will be discussed in the next section. In a dilute gas where all $\kappa_{\alpha\beta}>1$, the emission operators are simply $J_m\approx b_m$ and the decay rates are the single atom ones $\gamma_m\approx\gamma$ for all $m=1\dots N$, i.e. the photon emissions occur independently from each individual atom. However, as $\kappa_{\alpha\beta}$ decreases, the emission operators become superpositions of several atomic transition operators and the incoherent emission of photons occurs by means of collective \emph{superradiant} processes with $\gamma_{m}>\gamma$ and \emph{subradiant} ones with $\gamma_{m}<\gamma$. While the fraction of superradiant emission operators depends on the specific geometric arrangement and size of the system, it stays almost constant and small as the system size $N$ is increased. In the following, we will denote the largest collective decay rate $\gamma_\mathrm{S}$.

\begin{figure}[t]
\includegraphics[width=\columnwidth]{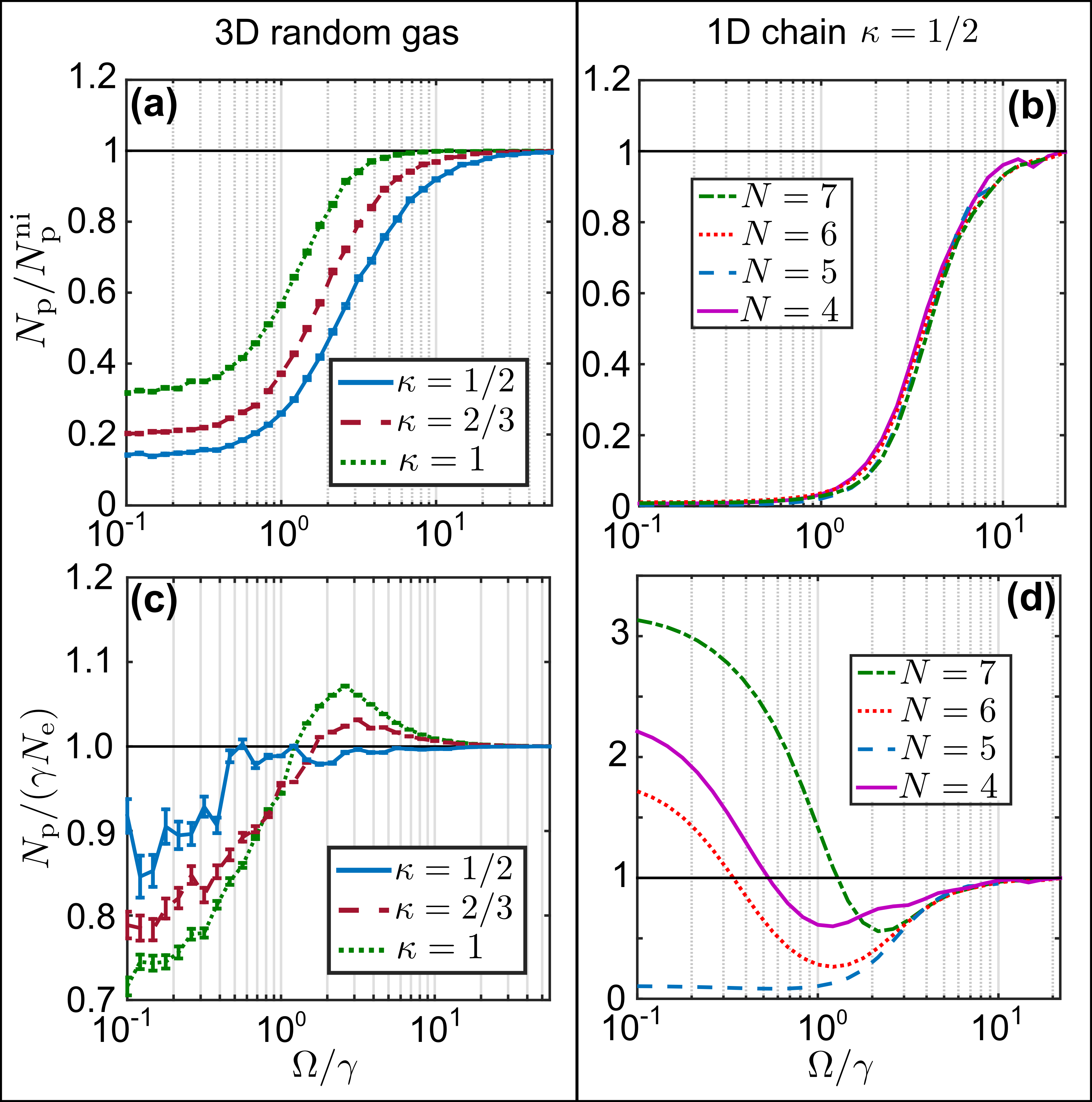}
\caption{(Color online) Photon emission rate in a dense gas $N_\mathrm{p}$ is suppressed with respect to a dilute one $N_\mathrm{p}^\mathrm{ni}$ for a large range of values of the driving $\Omega$. $N_\mathrm{p}/N_\mathrm{p}^\mathrm{ni}$ is shown \textbf{(a)} in a 3D random gas with $N=5$ and $\kappa=1,2/3$ and $1/2$ and \textbf{(b)} in a 1D chain with $\kappa=1/2$ and $N=4,5,6$ and 7. The emission rate is equal to the excitation number $N_\mathrm{e}$ times $\gamma$ in the dilute limit. We show $N_\mathrm{p}/(\gamma N_\mathrm{e})$ in a dense gas as a function of $\Omega/\gamma$ \textbf{(c)} in a 3D random gas with $N=5$ and \textbf{(d)} a 1D chain with $\kappa=1/2$.}\label{Fig3}
\end{figure}

\noindent\textit{Photon emission in ordered and disordered gases.} Our aim is to explore the signatures of collective behavior in the light scattered from the system in the stationary state. One of these signatures, which has been observed in experiments such as \cite{Pellegrino14} under weak driving conditions ($\Omega/\gamma\ll1$), is the suppression of the photon emission rate with respect to the dilute case. One can obtain this intensity in terms of the emission operators and rates discussed above as
\begin{equation}\label{eqn:Np}
  N_\mathrm{p}=\sum_{m=1}^N\gamma_m\left<J_m^\dag J_m\right>_\mathrm{ss},
\end{equation}
where $\left\langle\cdot\right\rangle_\mathrm{ss}$ denotes the expectation value in the stationary state. In Fig. \ref{Fig3}(a) we compare the results of the emission rate with the dilute (non-interacting) limit $N_\mathrm{p}^\mathrm{ni}$ for a 3D random gas, showing the average results of 1000 numerical experiments with different positions of the atoms. We enhance the collective behavior in the system by decreasing $\kappa=2\pi a/\lambda$, where $a$ represents the average distance between each atom and the one closest to it. We observe indeed strong suppression of the emission in the limit of weak driving, which is more pronounced the smaller $\kappa$ is. Beyond this limit, this suppression, although less pronounced, is still present for values of the driving $\Omega$ comparable to the single atom decay rate $\gamma$. Eventually, for large enough $\Omega/\gamma$, the suppression disappears and $N_\mathrm{p}/N_\mathrm{p}^\mathrm{ni}\to1$. This remains unchanged for the system sizes explored, which gives an indication that insights on the behavior of larger systems can be indeed extracted from these results. In order to explore the importance of the specific external geometry of the system, we calculate the ratio $N_\mathrm{p}/N_\mathrm{p}^\mathrm{ni}$ in a 1D chain perpendicular to the laser momentum with $\kappa=1/2$ [Fig. \ref{Fig3}(b)]. While the suppression is more pronounced in this case, we observe a very similar qualitative behavior to the disordered case. Again only minor differences exist between the results for $N=4,5,6$ and 7 atoms (shown in the figure). The results seem to indicate that the emission suppression is a very robust feature of these interacting systems with collective dissipation that survives the addition of finite driving and that does not depend on the specific spatial arrangement of the atoms.

As we discussed above, in a dilute gas the emission operators coincide with the atomic transition ones. Hence, here the emission rate (\ref{eqn:Np}) is equivalent to the number of excitations in the stationary state, $N_\mathrm{e}=\sum_{\alpha=1}^N\left<b_\alpha^\dag b_\alpha\right>_\mathrm{ss}$ multiplied by the single atom decay rate $\gamma$. This relation, however, does not hold in general as the dissipation acquires a collective character \cite{Ates12,Olmos14}. This feature can be observed in Figs. \ref{Fig3}(c) and (d), where the ratio between the photon emission rate and the excitation number multiplied by $\gamma$ is shown as a function of $\Omega/\gamma$. In the limit of very strong driving, both in the case of a 3D random gas and in a 1D chain this ratio approaches one. For smaller values of the driving, however, the geometry of the system becomes relevant. In the 1D chain [Fig. \ref{Fig3}(d)], one can observe finite size effects with the ratio $N_\mathrm{p}/(\gamma N_\mathrm{e})$ changing from larger to smaller than one as we vary the system size. Note [Fig. \ref{Fig3}(c)] that this feature is not present in the 3D random gas and $N_\mathrm{p}/(\gamma N_\mathrm{e})$ is largely independent of the system size. While this is true for the average over many configurations, the values for each independent configuration fluctuate notably from one to another, represented by large error bars. Moreover, we observe that as $\kappa$ decreases $N_\mathrm{p}/(\gamma N_\mathrm{e})$ approaches one.

\begin{figure}[t]
\includegraphics[width=\columnwidth]{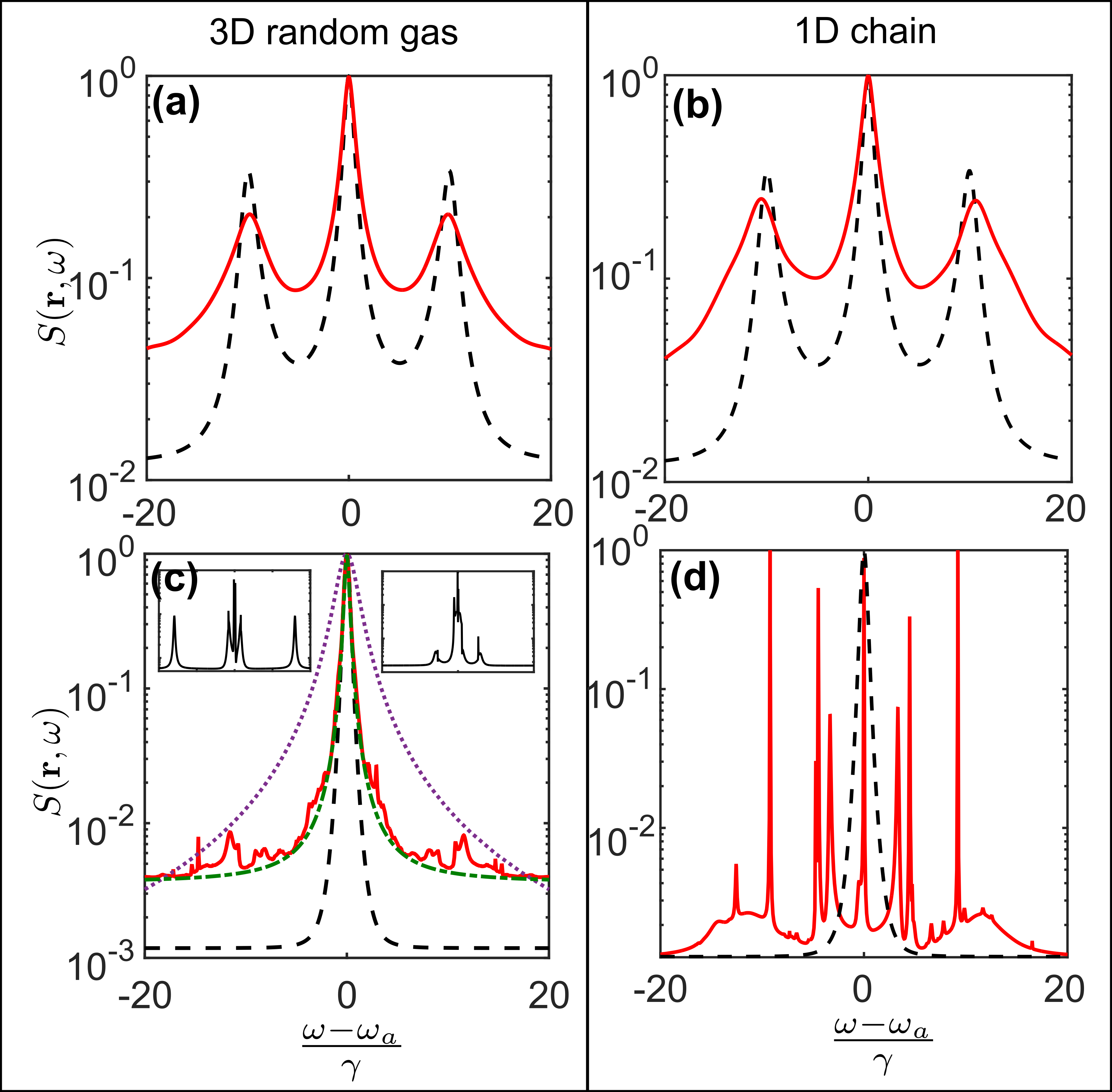}
\caption{(Color online) Power spectrum for a non-interacting (dashed black lines) and interacting (red solid lines) gas with $\kappa=1/2$ in: \textbf{(a)} and \textbf{(b)} show broadening of the Mollow triplet in the case of strong driving ($\Omega=10\gamma$) for a random 3D gas of $N=5$ atoms and 1D chain of $N=7$ atoms, respectively. \textbf{(c)} In the case of weak driving ($\Omega=0.1\gamma$) for a random 3D gas, the individual realizations give rise to spectra (insets) very different from the average one, which can be fitted with a Lorentzian (green dash-dotted line). For comparison, a Lorentzian with width $\gamma_\mathrm{S}$ is also shown (purple dotted line). \textbf{(d)} Spectrum of a 1D chain of $N=7$ atoms for $\Omega=0.1\gamma$.}\label{Fig4}
\end{figure}

\noindent\textit{Resonance Fluorescence.} Further signatures of collective behavior can be found in the spectral properties of the light emitted by the system. Here, we calculate numerically the power spectrum of the light emitted by the system in a fixed arbitrary position $\mathbf{r}=r\mathbf{\hat{r}}$ in the $xy$ plane \footnote{Note that only minor differences are found in the spectrum when changing the direction of observation within the $xy$ plane.}
\begin{equation*}
S(\mathbf{r},\omega)=\frac{1}{\pi}\textrm{Re}\int^{\infty}_{0}e^{i\omega\tau} \left<\mathbf{E}({\bf{r}},t)\mathbf{E}^\dag({\bf{r}},t+\tau)\right>_\mathrm{ss}d\tau,
\end{equation*}
where ${\bf{E}}({\bf{r}},t)$ denotes the negative-frequency part of the electric field operator in the far-zone approximation and in the Heisenberg picture, which is given by \cite{James93,BookScully08}
\begin{eqnarray*}
{\bf{E}}({\bf{r}},t)&=&\frac{\omega_\mathrm{a}^2}{4\pi c^2}\sum_{\alpha=1}^N\frac{\mathbf{\hat{r}}\times\mathbf{\hat{r}}\times\mathbf{d}}{\left|\mathbf{r}-\mathbf{r}_\alpha\right|} b^\dag_{\alpha}\left(t-\frac{|{\bf{r}}-{\bf{r}}_{\alpha}|}{c}\right),
\end{eqnarray*}
where $\omega_\mathrm{a}=2\pi c/\lambda$. The results of the numerical calculation of the power spectrum in the far field are shown in Fig.~\ref{Fig4} for two values of $\Omega/\gamma=0.1$ and $10$, representative of weak and strong driving, respectively. We once again compare the results for a 1D chain and a 3D random gas with $\kappa=1/2$ (averaged over 1000 realisations). In all cases we show the spectrum resulting from a dilute --non-interacting-- gas, which in the weak driving regime is formed by a single peak with width smaller than $\gamma$ and in the strong driving one by a so-called Mollow triplet \cite{Kimble76,BookScully08}.

When the system is strongly driven [Figs.~\ref{Fig4}(a) and (b)], the two geometries show similar features: the interactions in the system lead to a broadening of the three peaks of the Mollow triplet, with the sideband peaks appearing slightly shifted away from the central one in the 1D chain. In the weak driving case, however, the geometry of the system plays a rather important role. In the 1D chain we can observe a large number of peaks with width much larger (smaller) than the single atom one, corresponding to superradiant (subradiant) states [Fig.~\ref{Fig4}(d)]. Similar configurations with the peaks appearing in different positions are also visible in each individual realisation in the 3D random case [see insets in Fig. \ref{Fig4}(c)]. The averaging process over all these configurations, where the positions and amplitudes of the subradiant and superradiant peaks are shifted in every run gives as a result a single Lorentzian-like peak broader than the non-interacting one. Note that this width is noticeably smaller than the decay rate of the most superradiant mode $\gamma_\mathrm{S}$ \cite{Pellegrino14}. Finally, note that only negligible shifts of the central feature are observed in the data \cite{Jennewein15,Jenkins16}.

\noindent\textit{Second order correlation function.} The second-order correlation of the resonance fluorescence from different light sources has been widely investigated: It has been established that a thermal source emits photons in bunches, while antibunched photon emission is only seen in quantum light \cite{BookScully08,Diedrich87,Gerber09,Basche92}. We investigate here the second-order correlation function of the scattered light in our system, defined as
\begin{equation}\label{g2}
g^{(2)}(\tau)=\frac{\left\langle \mathbf{E}({\bf{r}},t)\mathbf{E}({\bf{r}},t+\tau)\mathbf{E}^\dag({\bf{r}},t+\tau)\mathbf{E}^\dag({\bf{r}},t)\right\rangle}{\left\langle I({\bf{r}},t)\right\rangle
\left\langle I({\bf{r}},t+\tau)\right\rangle},
\end{equation}
where $I({\bf{r}},t)=\mathbf{E}({\bf{r}},t)\mathbf{E}^\dag({\bf{r}},t)$. Equation~(\ref{g2}) yields the probability of detecting a photon at time $t+\tau$ given that one was detected at time $t$ divided by the probability of uncorrelated detection. As the atomic ensemble is in the stationary state, $t\rightarrow\infty$, the intensity correlation depends only on the time delay $\tau$.

\begin{figure}[t]
\includegraphics[width=\columnwidth]{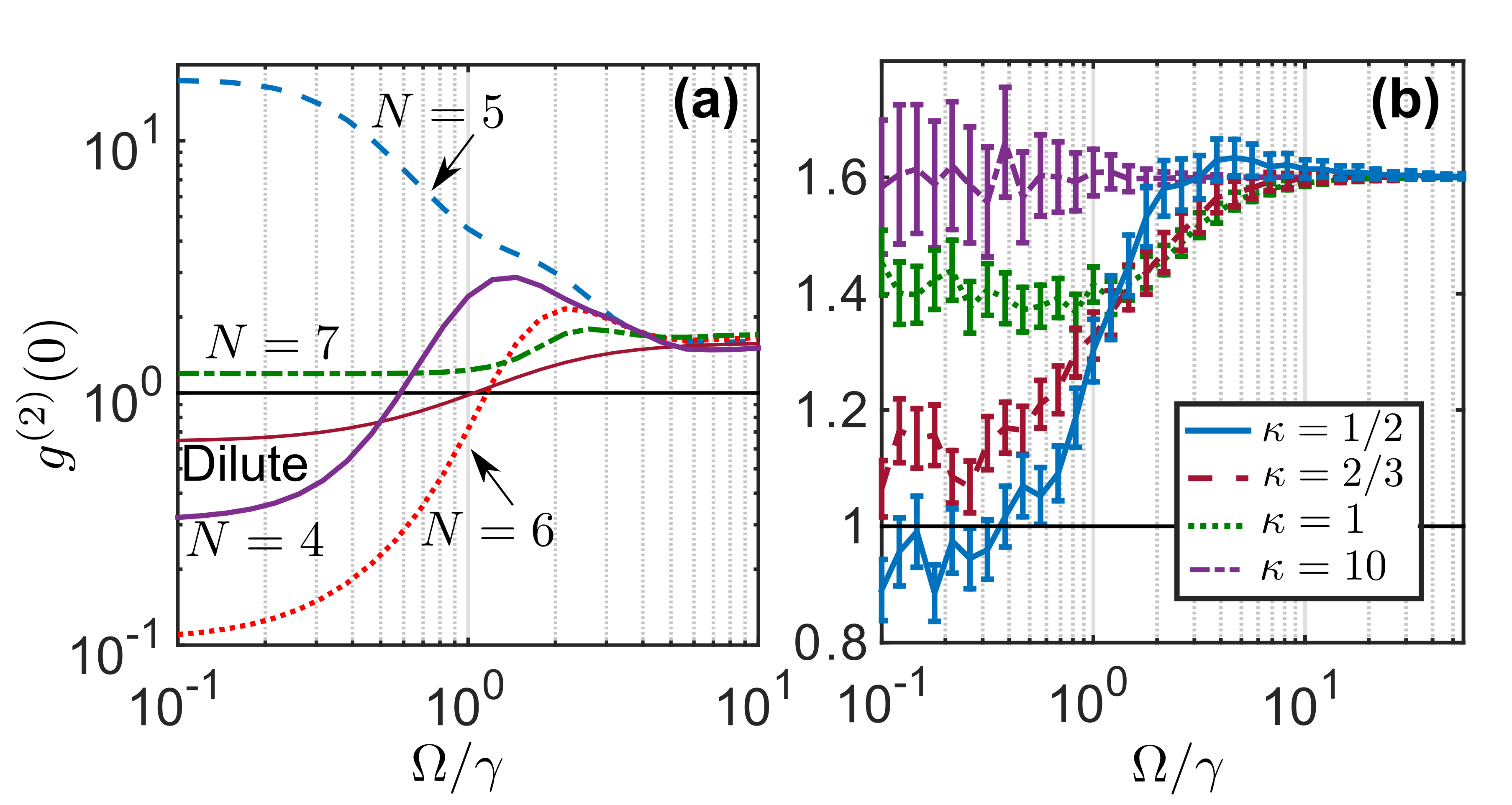}\\
\caption{(Color online) Second-order correlation function $g^{(2)}(\tau=0)$ as a function of $\Omega/\gamma$. \textbf{(a)} Data for a 1D chain with size $N=4,5,6$ and 7 with $\kappa=1/2$. \textbf{(b)} Data for a system of $N=5$ atoms in a 3D random gas for different values of $\kappa$. The dilute limit ($\kappa=10$) for $N=5$ is shown in both panels for comparison.}\label{Fig5}
\end{figure}

Figure~\ref{Fig5} displays data of the second-order correlation $g^{(2)}(\tau)$ at $\tau=0$ in a dense gas, where we show for comparison the results for a dilute one ($\kappa=10$). The geometry of the atomic ensemble has again a considerable impact on the form of the correlation function except in the limit of very strong driving, where its value tends to the same limit, $g^{(2)}(0)=2(1-1/N)$ \cite{Meiser10}. As it was the case for $N_\mathrm{p}/(\gamma N_\mathrm{e})$, $g^{(2)}(0)$ in the 1D chain shows strong finite size effects [Figure~\ref{Fig5}(a)]. In particular, when the number of atoms in the chain is odd one observes bunching ($g^{(2)}(0)>1$) while for even sizes the photon emission is antibunched \cite{Bhatti15,Auffeves11}. Conversely, in the 3D random gas [Figure~\ref{Fig5}(b)] $g^{(2)}(0)$ is largely independent of the system size. We observe that here the signature of the collective behavior in the system is a reduction of $g^{(2)}(0)$ with respect to the dilute case, even going from bunched to antibunched behavior for $\kappa=1/2$. This is particularly pronounced in the weak driving limit, where we also observe large fluctuations around the average \cite{Meiser10}.

\noindent\textit{Conclusions.} To conclude, let us return to the three questions posed in the introduction of the paper. (A) In the strong driving regime we observe suppression of photon emission rate, although less pronounced than the limit of weak driving. Moreover, clear signatures of the strong interactions and collective dissipation in this regime are visible in the broadening of the spectrum (Mollow triplet). (B) The fact that there is no simple one-to-one relation between the excitation density and the photon emission intensity can be observed in this system as a signature of collective behavior. (C) Finally, we show that in general the properties of the scattered light in a disordered gas are qualitatively different from the ones obtained from an ordered configuration. The reason can be found in the averaging process, that washes out the specific features of each single realisation.

We have shown that in a dense ordered gas the interaction effects are most pronounced. Experiments that explore the dense regime in these ordered configurations have not been performed yet. Strontium atoms possess a very long wavelength transition between low-lying levels and can be trapped in lattices with lattice constant on the order of a few hundred nm \cite{Olmos13}. Hence, they represent an ideal platform for the observation of collective effects in dense atomic gases.

\begin{acknowledgements}
\noindent\textit{Acknowledgements.} The authors would like to acknowledge Igor Lesanovsky for useful discussions. Also Michael R. Hush and Deshui Yu are acknowledged for discussions in the very early stages of this work. B.O acknowledges funding from the Royal Society. We are grateful for access to the University of Nottingham High Performance Computing Facility.
\end{acknowledgements}


\begin{thebibliography}{38}
\expandafter\ifx\csname natexlab\endcsname\relax\def\natexlab#1{#1}\fi
\expandafter\ifx\csname bibnamefont\endcsname\relax
  \def\bibnamefont#1{#1}\fi
\expandafter\ifx\csname bibfnamefont\endcsname\relax
  \def\bibfnamefont#1{#1}\fi
\expandafter\ifx\csname citenamefont\endcsname\relax
  \def\citenamefont#1{#1}\fi
\expandafter\ifx\csname url\endcsname\relax
  \def\url#1{\texttt{#1}}\fi
\expandafter\ifx\csname urlprefix\endcsname\relax\def\urlprefix{URL }\fi
\providecommand{\bibinfo}[2]{#2}
\providecommand{\eprint}[2][]{\url{#2}}

\bibitem[{\citenamefont{Dicke}(1954)}]{Dicke54}
\bibinfo{author}{\bibfnamefont{R.~H.} \bibnamefont{Dicke}},
  \bibinfo{journal}{Phys. Rev.} \textbf{\bibinfo{volume}{93}},
  \bibinfo{pages}{99} (\bibinfo{year}{1954}).

\bibitem[{\citenamefont{Lehmberg}(1970)}]{Lehmberg70}
\bibinfo{author}{\bibfnamefont{R.~H.} \bibnamefont{Lehmberg}},
  \bibinfo{journal}{Phys. Rev. A} \textbf{\bibinfo{volume}{2}},
  \bibinfo{pages}{883} (\bibinfo{year}{1970}).

\bibitem[{\citenamefont{Agarwal}(1970)}]{Agarwal70}
\bibinfo{author}{\bibfnamefont{G.~S.} \bibnamefont{Agarwal}},
  \bibinfo{journal}{Phys. Rev. A} \textbf{\bibinfo{volume}{2}},
  \bibinfo{pages}{2038} (\bibinfo{year}{1970}).

\bibitem[{\citenamefont{Bloch et~al.}(2008)\citenamefont{Bloch, Dalibard, and
  Zwerger}}]{Bloch08}
\bibinfo{author}{\bibfnamefont{I.}~\bibnamefont{Bloch}},
  \bibinfo{author}{\bibfnamefont{J.}~\bibnamefont{Dalibard}}, \bibnamefont{and}
  \bibinfo{author}{\bibfnamefont{W.}~\bibnamefont{Zwerger}},
  \bibinfo{journal}{Rev. Mod. Phys.} \textbf{\bibinfo{volume}{80}},
  \bibinfo{pages}{885} (\bibinfo{year}{2008}).

\bibitem[{\citenamefont{R\"ohlsberger et~al.}(2010)\citenamefont{R\"ohlsberger,
  Schlage, Sahoo, Couet, and R\"uffer}}]{Rohlsberger10}
\bibinfo{author}{\bibfnamefont{R.}~\bibnamefont{R\"ohlsberger}},
  \bibinfo{author}{\bibfnamefont{K.}~\bibnamefont{Schlage}},
  \bibinfo{author}{\bibfnamefont{B.}~\bibnamefont{Sahoo}},
  \bibinfo{author}{\bibfnamefont{S.}~\bibnamefont{Couet}}, \bibnamefont{and}
  \bibinfo{author}{\bibfnamefont{R.}~\bibnamefont{R\"uffer}},
  \bibinfo{journal}{Science} \textbf{\bibinfo{volume}{328}},
  \bibinfo{pages}{1248} (\bibinfo{year}{2010}).

\bibitem[{\citenamefont{Keaveney et~al.}(2012)\citenamefont{Keaveney, Sargsyan,
  Krohn, Hughes, Sarkisyan, and Adams}}]{Keaveney12}
\bibinfo{author}{\bibfnamefont{J.}~\bibnamefont{Keaveney}},
  \bibinfo{author}{\bibfnamefont{A.}~\bibnamefont{Sargsyan}},
  \bibinfo{author}{\bibfnamefont{U.}~\bibnamefont{Krohn}},
  \bibinfo{author}{\bibfnamefont{I.~G.} \bibnamefont{Hughes}},
  \bibinfo{author}{\bibfnamefont{D.}~\bibnamefont{Sarkisyan}},
  \bibnamefont{and} \bibinfo{author}{\bibfnamefont{C.~S.} \bibnamefont{Adams}},
  \bibinfo{journal}{Phys. Rev. Lett.} \textbf{\bibinfo{volume}{108}},
  \bibinfo{pages}{173601} (\bibinfo{year}{2012}).

\bibitem[{\citenamefont{Meir et~al.}(2014)\citenamefont{Meir, Schwartz,
  Shahmoon, Oron, and Ozeri}}]{Meir14}
\bibinfo{author}{\bibfnamefont{Z.}~\bibnamefont{Meir}},
  \bibinfo{author}{\bibfnamefont{O.}~\bibnamefont{Schwartz}},
  \bibinfo{author}{\bibfnamefont{E.}~\bibnamefont{Shahmoon}},
  \bibinfo{author}{\bibfnamefont{D.}~\bibnamefont{Oron}}, \bibnamefont{and}
  \bibinfo{author}{\bibfnamefont{R.}~\bibnamefont{Ozeri}},
  \bibinfo{journal}{Phys. Rev. Lett.} \textbf{\bibinfo{volume}{113}},
  \bibinfo{pages}{193002} (\bibinfo{year}{2014}).

\bibitem[{\citenamefont{Pellegrino et~al.}(2014)\citenamefont{Pellegrino,
  Bourgain, Jennewein, Sortais, Browaeys, Jenkins, and
  Ruostekoski}}]{Pellegrino14}
\bibinfo{author}{\bibfnamefont{J.}~\bibnamefont{Pellegrino}},
  \bibinfo{author}{\bibfnamefont{R.}~\bibnamefont{Bourgain}},
  \bibinfo{author}{\bibfnamefont{S.}~\bibnamefont{Jennewein}},
  \bibinfo{author}{\bibfnamefont{Y.~R.~P.} \bibnamefont{Sortais}},
  \bibinfo{author}{\bibfnamefont{A.}~\bibnamefont{Browaeys}},
  \bibinfo{author}{\bibfnamefont{S.~D.} \bibnamefont{Jenkins}},
  \bibnamefont{and}
  \bibinfo{author}{\bibfnamefont{J.}~\bibnamefont{Ruostekoski}},
  \bibinfo{journal}{Phys. Rev. Lett.} \textbf{\bibinfo{volume}{113}},
  \bibinfo{pages}{133602} (\bibinfo{year}{2014}).

\bibitem[{\citenamefont{Kwong et~al.}(2014)\citenamefont{Kwong, Yang, Pramod,
  Pandey, Delande, Pierrat, and Wilkowski}}]{Kwong14}
\bibinfo{author}{\bibfnamefont{C.~C.} \bibnamefont{Kwong}},
  \bibinfo{author}{\bibfnamefont{T.}~\bibnamefont{Yang}},
  \bibinfo{author}{\bibfnamefont{M.~S.} \bibnamefont{Pramod}},
  \bibinfo{author}{\bibfnamefont{K.}~\bibnamefont{Pandey}},
  \bibinfo{author}{\bibfnamefont{D.}~\bibnamefont{Delande}},
  \bibinfo{author}{\bibfnamefont{R.}~\bibnamefont{Pierrat}}, \bibnamefont{and}
  \bibinfo{author}{\bibfnamefont{D.}~\bibnamefont{Wilkowski}},
  \bibinfo{journal}{Phys. Rev. Lett.} \textbf{\bibinfo{volume}{113}},
  \bibinfo{pages}{223601} (\bibinfo{year}{2014}).

\bibitem[{\citenamefont{Kwong et~al.}(2015)\citenamefont{Kwong, Yang, Delande,
  Pierrat, and Wilkowski}}]{Kwong15}
\bibinfo{author}{\bibfnamefont{C.~C.} \bibnamefont{Kwong}},
  \bibinfo{author}{\bibfnamefont{T.}~\bibnamefont{Yang}},
  \bibinfo{author}{\bibfnamefont{D.}~\bibnamefont{Delande}},
  \bibinfo{author}{\bibfnamefont{R.}~\bibnamefont{Pierrat}}, \bibnamefont{and}
  \bibinfo{author}{\bibfnamefont{D.}~\bibnamefont{Wilkowski}},
  \bibinfo{journal}{Phys. Rev. Lett.} \textbf{\bibinfo{volume}{115}},
  \bibinfo{pages}{223601} (\bibinfo{year}{2015}).

\bibitem[{\citenamefont{Jennewein et~al.}(2015)\citenamefont{Jennewein, Besbes,
  Schilder, Jenkins, Sauvan, Ruostekoski, Greffet, Sortais, and
  Browaeys}}]{Jennewein15}
\bibinfo{author}{\bibfnamefont{S.}~\bibnamefont{Jennewein}},
  \bibinfo{author}{\bibfnamefont{M.}~\bibnamefont{Besbes}},
  \bibinfo{author}{\bibfnamefont{N.}~\bibnamefont{Schilder}},
  \bibinfo{author}{\bibfnamefont{S.}~\bibnamefont{Jenkins}},
  \bibinfo{author}{\bibfnamefont{C.}~\bibnamefont{Sauvan}},
  \bibinfo{author}{\bibfnamefont{J.}~\bibnamefont{Ruostekoski}},
  \bibinfo{author}{\bibfnamefont{J.-J.} \bibnamefont{Greffet}},
  \bibinfo{author}{\bibfnamefont{Y.}~\bibnamefont{Sortais}}, \bibnamefont{and}
  \bibinfo{author}{\bibfnamefont{A.}~\bibnamefont{Browaeys}},
  \bibinfo{journal}{preprint} p. \bibinfo{pages}{arXiv:1510.08041}
  (\bibinfo{year}{2015}).

\bibitem[{\citenamefont{Bromley et~al.}(2016)\citenamefont{Bromley, Zhu,
  Bishof, Zhang, Bothwell, Schachenmayer, Nicholson, Kaiser, Yelin, Lukin
  et~al.}}]{Bromley16}
\bibinfo{author}{\bibfnamefont{S.~L.} \bibnamefont{Bromley}},
  \bibinfo{author}{\bibfnamefont{B.}~\bibnamefont{Zhu}},
  \bibinfo{author}{\bibfnamefont{M.}~\bibnamefont{Bishof}},
  \bibinfo{author}{\bibfnamefont{X.}~\bibnamefont{Zhang}},
  \bibinfo{author}{\bibfnamefont{T.}~\bibnamefont{Bothwell}},
  \bibinfo{author}{\bibfnamefont{J.}~\bibnamefont{Schachenmayer}},
  \bibinfo{author}{\bibfnamefont{T.~L.} \bibnamefont{Nicholson}},
  \bibinfo{author}{\bibfnamefont{R.}~\bibnamefont{Kaiser}},
  \bibinfo{author}{\bibfnamefont{S.~F.} \bibnamefont{Yelin}},
  \bibinfo{author}{\bibfnamefont{M.~D.} \bibnamefont{Lukin}},
  \bibnamefont{et~al.}, \bibinfo{journal}{preprint} p.
  \bibinfo{pages}{arXiv:1601.05322} (\bibinfo{year}{2016}).

\bibitem[{\citenamefont{Jenkins et~al.}(2016)\citenamefont{Jenkins,
  Ruostekoski, Javanainen, Bourgain, Jennewein, Sortais, and
  Browaeys}}]{Jenkins16}
\bibinfo{author}{\bibfnamefont{S.~D.} \bibnamefont{Jenkins}},
  \bibinfo{author}{\bibfnamefont{J.}~\bibnamefont{Ruostekoski}},
  \bibinfo{author}{\bibfnamefont{J.}~\bibnamefont{Javanainen}},
  \bibinfo{author}{\bibfnamefont{R.}~\bibnamefont{Bourgain}},
  \bibinfo{author}{\bibfnamefont{S.}~\bibnamefont{Jennewein}},
  \bibinfo{author}{\bibfnamefont{Y.~R.~P.} \bibnamefont{Sortais}},
  \bibnamefont{and} \bibinfo{author}{\bibfnamefont{A.}~\bibnamefont{Browaeys}},
  \bibinfo{journal}{preprint} p. \bibinfo{pages}{arXiv:1602.01037}
  (\bibinfo{year}{2016}).

\bibitem[{\citenamefont{Fleischhauer and Yelin}(1999)}]{Fleischhauer99}
\bibinfo{author}{\bibfnamefont{M.}~\bibnamefont{Fleischhauer}}
  \bibnamefont{and} \bibinfo{author}{\bibfnamefont{S.~F.} \bibnamefont{Yelin}},
  \bibinfo{journal}{Phys. Rev. A} \textbf{\bibinfo{volume}{59}},
  \bibinfo{pages}{2427} (\bibinfo{year}{1999}).

\bibitem[{\citenamefont{Scully}(2009)}]{Scully09}
\bibinfo{author}{\bibfnamefont{M.~O.} \bibnamefont{Scully}},
  \bibinfo{journal}{Phys. Rev. Lett.} \textbf{\bibinfo{volume}{102}},
  \bibinfo{pages}{143601} (\bibinfo{year}{2009}).

\bibitem[{\citenamefont{Svidzinsky et~al.}(2010)\citenamefont{Svidzinsky,
  Chang, and Scully}}]{Svidzinsky10}
\bibinfo{author}{\bibfnamefont{A.~A.} \bibnamefont{Svidzinsky}},
  \bibinfo{author}{\bibfnamefont{J.-T.} \bibnamefont{Chang}}, \bibnamefont{and}
  \bibinfo{author}{\bibfnamefont{M.~O.} \bibnamefont{Scully}},
  \bibinfo{journal}{Phys. Rev. A} \textbf{\bibinfo{volume}{81}},
  \bibinfo{pages}{053821} (\bibinfo{year}{2010}).

\bibitem[{\citenamefont{Bienaim\'e et~al.}(2012)\citenamefont{Bienaim\'e,
  Piovella, and Kaiser}}]{Bienaime12}
\bibinfo{author}{\bibfnamefont{T.}~\bibnamefont{Bienaim\'e}},
  \bibinfo{author}{\bibfnamefont{N.}~\bibnamefont{Piovella}}, \bibnamefont{and}
  \bibinfo{author}{\bibfnamefont{R.}~\bibnamefont{Kaiser}},
  \bibinfo{journal}{Phys. Rev. Lett.} \textbf{\bibinfo{volume}{108}},
  \bibinfo{pages}{123602} (\bibinfo{year}{2012}).

\bibitem[{\citenamefont{Bienaim\'{e} et~al.}(2013)\citenamefont{Bienaim\'{e},
  Bachelard, Piovella, and Kaiser}}]{Bienaime13}
\bibinfo{author}{\bibfnamefont{T.}~\bibnamefont{Bienaim\'{e}}},
  \bibinfo{author}{\bibfnamefont{R.}~\bibnamefont{Bachelard}},
  \bibinfo{author}{\bibfnamefont{N.}~\bibnamefont{Piovella}}, \bibnamefont{and}
  \bibinfo{author}{\bibfnamefont{R.}~\bibnamefont{Kaiser}},
  \bibinfo{journal}{Fortschr. Phys.} \textbf{\bibinfo{volume}{61}},
  \bibinfo{pages}{377} (\bibinfo{year}{2013}).

\bibitem[{\citenamefont{Olmos et~al.}(2013)\citenamefont{Olmos, Yu, Singh,
  Schreck, Bongs, and Lesanovsky}}]{Olmos13}
\bibinfo{author}{\bibfnamefont{B.}~\bibnamefont{Olmos}},
  \bibinfo{author}{\bibfnamefont{D.}~\bibnamefont{Yu}},
  \bibinfo{author}{\bibfnamefont{Y.}~\bibnamefont{Singh}},
  \bibinfo{author}{\bibfnamefont{F.}~\bibnamefont{Schreck}},
  \bibinfo{author}{\bibfnamefont{K.}~\bibnamefont{Bongs}}, \bibnamefont{and}
  \bibinfo{author}{\bibfnamefont{I.}~\bibnamefont{Lesanovsky}},
  \bibinfo{journal}{Phys. Rev. Lett.} \textbf{\bibinfo{volume}{110}},
  \bibinfo{pages}{143602} (\bibinfo{year}{2013}).

\bibitem[{\citenamefont{Li et~al.}(2013)\citenamefont{Li, Evers, Feng, and
  Zhu}}]{Li13}
\bibinfo{author}{\bibfnamefont{Y.}~\bibnamefont{Li}},
  \bibinfo{author}{\bibfnamefont{J.}~\bibnamefont{Evers}},
  \bibinfo{author}{\bibfnamefont{W.}~\bibnamefont{Feng}}, \bibnamefont{and}
  \bibinfo{author}{\bibfnamefont{S.-Y.} \bibnamefont{Zhu}},
  \bibinfo{journal}{Phys. Rev. A} \textbf{\bibinfo{volume}{87}},
  \bibinfo{pages}{053837} (\bibinfo{year}{2013}).

\bibitem[{\citenamefont{Bettles et~al.}(2015)\citenamefont{Bettles, Gardiner,
  and Adams}}]{Bettles15}
\bibinfo{author}{\bibfnamefont{R.~J.} \bibnamefont{Bettles}},
  \bibinfo{author}{\bibfnamefont{S.~A.} \bibnamefont{Gardiner}},
  \bibnamefont{and} \bibinfo{author}{\bibfnamefont{C.~S.} \bibnamefont{Adams}},
  \bibinfo{journal}{Phys. Rev. A} \textbf{\bibinfo{volume}{92}},
  \bibinfo{pages}{063822} (\bibinfo{year}{2015}).

\bibitem[{\citenamefont{Ku\ifmmode~\acute{s}\else \'{s}\fi{} and
  W\'odkiewicz}(1981)}]{Kus81}
\bibinfo{author}{\bibfnamefont{M.}~\bibnamefont{Ku\ifmmode~\acute{s}\else
  \'{s}\fi{}}} \bibnamefont{and}
  \bibinfo{author}{\bibfnamefont{K.}~\bibnamefont{W\'odkiewicz}},
  \bibinfo{journal}{Phys. Rev. A} \textbf{\bibinfo{volume}{23}},
  \bibinfo{pages}{853} (\bibinfo{year}{1981}).

\bibitem[{\citenamefont{James}(1993)}]{James93}
\bibinfo{author}{\bibfnamefont{D.~F.~V.} \bibnamefont{James}},
  \bibinfo{journal}{Phys. Rev. A} \textbf{\bibinfo{volume}{47}},
  \bibinfo{pages}{1336} (\bibinfo{year}{1993}).

\bibitem[{\citenamefont{Hettich et~al.}(2002)\citenamefont{Hettich, Schmitt,
  Zitzmann, K\"u{}hn, Gerhardt, and Sandoghdar}}]{Hettich02}
\bibinfo{author}{\bibfnamefont{C.}~\bibnamefont{Hettich}},
  \bibinfo{author}{\bibfnamefont{C.}~\bibnamefont{Schmitt}},
  \bibinfo{author}{\bibfnamefont{J.}~\bibnamefont{Zitzmann}},
  \bibinfo{author}{\bibfnamefont{S.}~\bibnamefont{K\"u{}hn}},
  \bibinfo{author}{\bibfnamefont{I.}~\bibnamefont{Gerhardt}}, \bibnamefont{and}
  \bibinfo{author}{\bibfnamefont{V.}~\bibnamefont{Sandoghdar}},
  \bibinfo{journal}{Science} \textbf{\bibinfo{volume}{298}},
  \bibinfo{pages}{385} (\bibinfo{year}{2002}).

\bibitem[{\citenamefont{Das et~al.}(2008)\citenamefont{Das, Agarwal, and
  Scully}}]{Das08}
\bibinfo{author}{\bibfnamefont{S.}~\bibnamefont{Das}},
  \bibinfo{author}{\bibfnamefont{G.~S.} \bibnamefont{Agarwal}},
  \bibnamefont{and} \bibinfo{author}{\bibfnamefont{M.~O.}
  \bibnamefont{Scully}}, \bibinfo{journal}{Phys. Rev. Lett.}
  \textbf{\bibinfo{volume}{101}}, \bibinfo{pages}{153601}
  (\bibinfo{year}{2008}).

\bibitem[{\citenamefont{Wang et~al.}(2010)\citenamefont{Wang, Li, Zheng, and
  Zhu}}]{Wang10}
\bibinfo{author}{\bibfnamefont{D.-w.} \bibnamefont{Wang}},
  \bibinfo{author}{\bibfnamefont{Z.-h.} \bibnamefont{Li}},
  \bibinfo{author}{\bibfnamefont{H.}~\bibnamefont{Zheng}}, \bibnamefont{and}
  \bibinfo{author}{\bibfnamefont{S.-y.} \bibnamefont{Zhu}},
  \bibinfo{journal}{Phys. Rev. A} \textbf{\bibinfo{volume}{81}},
  \bibinfo{pages}{043819} (\bibinfo{year}{2010}).

\bibitem[{\citenamefont{Zoubi and Ritsch}(2012)}]{Zoubi12}
\bibinfo{author}{\bibfnamefont{H.}~\bibnamefont{Zoubi}} \bibnamefont{and}
  \bibinfo{author}{\bibfnamefont{H.}~\bibnamefont{Ritsch}},
  \bibinfo{journal}{Eur. Phys. J. D} \textbf{\bibinfo{volume}{66}},
  \bibinfo{pages}{292} (\bibinfo{year}{2012}).

\bibitem[{\citenamefont{Ott et~al.}(2013)\citenamefont{Ott, Wubs, Lodahl,
  Mortensen, and Kaiser}}]{Ott13}
\bibinfo{author}{\bibfnamefont{J.~R.} \bibnamefont{Ott}},
  \bibinfo{author}{\bibfnamefont{M.}~\bibnamefont{Wubs}},
  \bibinfo{author}{\bibfnamefont{P.}~\bibnamefont{Lodahl}},
  \bibinfo{author}{\bibfnamefont{N.~A.} \bibnamefont{Mortensen}},
  \bibnamefont{and} \bibinfo{author}{\bibfnamefont{R.}~\bibnamefont{Kaiser}},
  \bibinfo{journal}{Phys. Rev. A} \textbf{\bibinfo{volume}{87}},
  \bibinfo{pages}{061801} (\bibinfo{year}{2013}).

\bibitem[{\citenamefont{Ates et~al.}(2012)\citenamefont{Ates, Olmos, Garrahan,
  and Lesanovsky}}]{Ates12}
\bibinfo{author}{\bibfnamefont{C.}~\bibnamefont{Ates}},
  \bibinfo{author}{\bibfnamefont{B.}~\bibnamefont{Olmos}},
  \bibinfo{author}{\bibfnamefont{J.~P.} \bibnamefont{Garrahan}},
  \bibnamefont{and}
  \bibinfo{author}{\bibfnamefont{I.}~\bibnamefont{Lesanovsky}},
  \bibinfo{journal}{Phys. Rev. A} \textbf{\bibinfo{volume}{85}},
  \bibinfo{pages}{043620} (\bibinfo{year}{2012}).

\bibitem[{\citenamefont{Olmos et~al.}(2014)\citenamefont{Olmos, Yu, and
  Lesanovsky}}]{Olmos14}
\bibinfo{author}{\bibfnamefont{B.}~\bibnamefont{Olmos}},
  \bibinfo{author}{\bibfnamefont{D.}~\bibnamefont{Yu}}, \bibnamefont{and}
  \bibinfo{author}{\bibfnamefont{I.}~\bibnamefont{Lesanovsky}},
  \bibinfo{journal}{Phys. Rev. A} \textbf{\bibinfo{volume}{89}},
  \bibinfo{pages}{023616} (\bibinfo{year}{2014}).

\bibitem[{\citenamefont{Scully and Zubairy}(2008)}]{BookScully08}
\bibinfo{author}{\bibfnamefont{M.}~\bibnamefont{Scully}} \bibnamefont{and}
  \bibinfo{author}{\bibfnamefont{M.}~\bibnamefont{Zubairy}},
  \emph{\bibinfo{title}{Quantum Optics}} (\bibinfo{publisher}{Cambridge
  University Press}, \bibinfo{address}{Cambridge}, \bibinfo{year}{2008}),
  \bibinfo{edition}{6th} ed.

\bibitem[{\citenamefont{Kimble and Mandel}(1976)}]{Kimble76}
\bibinfo{author}{\bibfnamefont{H.~J.} \bibnamefont{Kimble}} \bibnamefont{and}
  \bibinfo{author}{\bibfnamefont{L.}~\bibnamefont{Mandel}},
  \bibinfo{journal}{Phys. Rev. A} \textbf{\bibinfo{volume}{13}},
  \bibinfo{pages}{2123} (\bibinfo{year}{1976}).

\bibitem[{\citenamefont{Diedrich and Walther}(1987)}]{Diedrich87}
\bibinfo{author}{\bibfnamefont{F.}~\bibnamefont{Diedrich}} \bibnamefont{and}
  \bibinfo{author}{\bibfnamefont{H.}~\bibnamefont{Walther}},
  \bibinfo{journal}{Phys. Rev. Lett.} \textbf{\bibinfo{volume}{58}},
  \bibinfo{pages}{203} (\bibinfo{year}{1987}).

\bibitem[{\citenamefont{Gerber et~al.}(2009)\citenamefont{Gerber, Rotter,
  Slodi\ifmmode~\check{c}\else \v{c}\fi{}ka, Eschner, Carmichael, and
  Blatt}}]{Gerber09}
\bibinfo{author}{\bibfnamefont{S.}~\bibnamefont{Gerber}},
  \bibinfo{author}{\bibfnamefont{D.}~\bibnamefont{Rotter}},
  \bibinfo{author}{\bibfnamefont{L.}~\bibnamefont{Slodi\ifmmode~\check{c}\else
  \v{c}\fi{}ka}}, \bibinfo{author}{\bibfnamefont{J.}~\bibnamefont{Eschner}},
  \bibinfo{author}{\bibfnamefont{H.~J.} \bibnamefont{Carmichael}},
  \bibnamefont{and} \bibinfo{author}{\bibfnamefont{R.}~\bibnamefont{Blatt}},
  \bibinfo{journal}{Phys. Rev. Lett.} \textbf{\bibinfo{volume}{102}},
  \bibinfo{pages}{183601} (\bibinfo{year}{2009}).

\bibitem[{\citenamefont{Basch\'e et~al.}(1992)\citenamefont{Basch\'e, Moerner,
  Orrit, and Talon}}]{Basche92}
\bibinfo{author}{\bibfnamefont{T.}~\bibnamefont{Basch\'e}},
  \bibinfo{author}{\bibfnamefont{W.~E.} \bibnamefont{Moerner}},
  \bibinfo{author}{\bibfnamefont{M.}~\bibnamefont{Orrit}}, \bibnamefont{and}
  \bibinfo{author}{\bibfnamefont{H.}~\bibnamefont{Talon}},
  \bibinfo{journal}{Phys. Rev. Lett.} \textbf{\bibinfo{volume}{69}},
  \bibinfo{pages}{1516} (\bibinfo{year}{1992}).

\bibitem[{\citenamefont{Meiser and Holland}(2010)}]{Meiser10}
\bibinfo{author}{\bibfnamefont{D.}~\bibnamefont{Meiser}} \bibnamefont{and}
  \bibinfo{author}{\bibfnamefont{M.~J.} \bibnamefont{Holland}},
  \bibinfo{journal}{Phys. Rev. A} \textbf{\bibinfo{volume}{81}},
  \bibinfo{pages}{063827} (\bibinfo{year}{2010}).

\bibitem[{\citenamefont{Bhatti et~al.}(2015)\citenamefont{Bhatti, von Zanthier,
  and Agarwal}}]{Bhatti15}
\bibinfo{author}{\bibfnamefont{D.}~\bibnamefont{Bhatti}},
  \bibinfo{author}{\bibfnamefont{J.}~\bibnamefont{von Zanthier}},
  \bibnamefont{and} \bibinfo{author}{\bibfnamefont{G.}~\bibnamefont{Agarwal}},
  \bibinfo{journal}{Sci. Rep.} \textbf{\bibinfo{volume}{5}},
  \bibinfo{pages}{17335} (\bibinfo{year}{2015}).

\bibitem[{\citenamefont{Auffèves et~al.}(2011)\citenamefont{Auffèves, Gerace,
  Portolan, Drezet, and França~Santos}}]{Auffeves11}
\bibinfo{author}{\bibfnamefont{A.}~\bibnamefont{Auffèves}},
  \bibinfo{author}{\bibfnamefont{D.}~\bibnamefont{Gerace}},
  \bibinfo{author}{\bibfnamefont{S.}~\bibnamefont{Portolan}},
  \bibinfo{author}{\bibfnamefont{A.}~\bibnamefont{Drezet}}, \bibnamefont{and}
  \bibinfo{author}{\bibfnamefont{M.}~\bibnamefont{França~Santos}},
  \bibinfo{journal}{New Journal of Physics} \textbf{\bibinfo{volume}{13}},
  \bibinfo{pages}{093020} (\bibinfo{year}{2011}).

\end{thebibliography}
\end{document}